# Overcoming Limitations in Artificial Intelligence based Prostate Cancer Detection with Improved Datasets and a Bayesian Framework for Panel Predictions

T. J. Hart[1], Chloe Engler Hart[1], Spencer Hopson[1], Paul M. Urie[1], Dennis Della Corte [1,*]

[1]Department of Physics and Astronomy, Brigham Young University, Provo.

*Dennis.DellaCorte@byu.edu

**Abstract**

Despite considerable progress in developing artificial intelligence (AI) algorithms for prostate cancer detection from whole slide images, the clinical applicability of these models remains limited due to variability in pathological annotations and existing dataset limitations. This article proposes a novel approach to overcome these challenges by leveraging a Bayesian framework to seamlessly integrate new data, and present results as a panel of annotations. The framework is demonstrated by integrating a Bayesian prior with one trained AI model to generate a distribution of Gleason patterns for each pixel of an image. It is shown that using this distribution of Gleason patterns rather than a ground-truth label can improve model applicability, mitigate errors, and highlight areas of interest for pathologists. Additionally, we present a high-quality, hand-curated dataset of prostate histopathological images annotated at the gland level by trained pre-medical students and verified by an expert pathologist. We highlight the potential of this adaptive and uncertainty-aware framework for developing clinically deployable AI tools that can support pathologists in accurate prostate cancer grading, improve diagnostic accuracy, and create positive patient outcomes.

**Introduction**

Prostate cancer is one of the most prevalent cancers in men, with the CDC (Centers for Disease Control) reporting 201,082 cases and 32,707 deaths in 2020 [1]. Crucial to early detection and accurate diagnosis, manually assigning Gleason grades and an overall Gleason score to whole slide images remains the standard for evaluating prostate cancer presence and severity [2]. Recently, artificial intelligence (AI) algorithms have shown promise in addressing the challenges of prostate cancer diagnosis, such as reducing diagnostic variability, increasing efficiency, and serving as a safeguard against misdiagnosis [3,4,5,6,7,8,9]. However, effective clinical implementation of AI tools for reliable Gleason grading remains elusive due to several key limitations.

First, the subjectivity inherent in the Gleason grading system, which ranks glandular tissue patterns on a severity scale of 1-5, leads to significant inter- and intra-observer variability among pathologists [10,11]. This poses a challenge for AI algorithms trained to predict a single ground truth label, as the algorithm will only perform as well as the grader on which it is trained and cannot account for the variability between gradings. Accounting for this variability is essential to assigning correct treatment and attaining the optimal prognosis since treatment depends upon Gleason scores [2].

Secondly, current public pathology datasets, such as the PANDA dataset, suffer from limitations that detract from overall accuracy and generalizability [10,13]. These limitations include annotation accuracy and quantity, sample diversity, realistic representations of cancer patterns, and a lack of diverse physician opinions. These shortcomings hinder the generalizability of AI models trained on these datasets and do not accurately represent a panel of opinions that accounts for observer variability [10,11,12].

To overcome these challenges, we introduce a new hand-curated dataset and propose a model framework leveraging Bayesian inference [14]. By curating a high-quality dataset of prostate histopathological images, we aim to introduce a method that captures the diversity of expert opinions. This dataset can be used to train AI models that predict distributions of Gleason patterns rather than a single ground truth label. Furthermore, we introduce a Bayesian framework that can combine the outputs of multiple AI models and opinions, produce results akin to a panel of experts, and update prior distributions as new models or data become available [14, 15, 16]. This method provides a more clinically relevant and generalizable prediction process for prostate cancer detection and grading automation.

## Methods

*Gland-Level Annotation Dataset Curation.* To create the dataset, a team of pre-medical students received rigorous training from an experienced pathologist. Using the QuPath digital pathology platform, the students annotated individual glands in prostate histopathology whole slide images, delineating their boundaries and assigning Gleason pattern and tissue border labels (Figure 1). The annotation process follows a hierarchical structure, with initial quality reviews conducted by senior students, followed by a final verification by the lead pathologist.

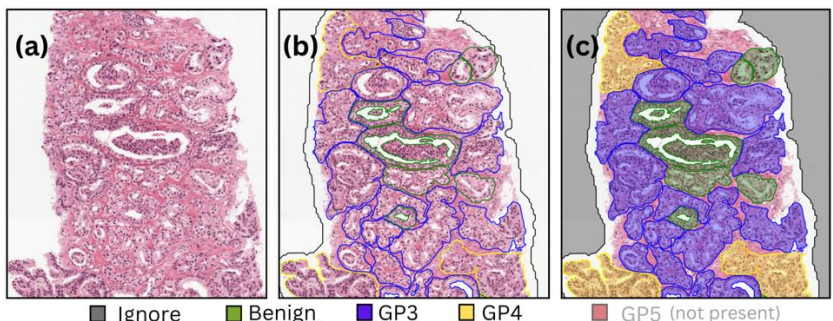

**Figure 1**. Prostate WSIs are annotated on a gland level and assigned Gleason Patterns. (a) Section of an unannotated prostate whole slide image. (b) Completed and graded annotations (unfilled). (c) Completed and graded annotations (filled).

*Hierarchical Bayesian modeling.* Hierarchical Bayesian modeling combines predictions from multiple AI models and captures the inherent uncertainty in Gleason pattern annotations from expert pathologists [14,15, 16]. Let $P = \{p_1, p_2, \ldots, p_n\}$ be the set of pixels in each prostate histopathology image and $G = \{g_1, g_2, \ldots, g_k\}$ be the set of possible Gleason patterns (e.g., Ignore, Tissue, Benign, GP3, GP4, GP5). Our goal is to estimate the posterior probability distribution $P(g_i | p_j, image)$ of Gleason patterns $g_i$ for each pixel $p_j$ given the other pixels in the image (denoted as image).

*Bayesian Inference Framework.* For the first AI model, the framework is implemented as follows:
1) Prior Distribution: The prior distribution $P(g_i | image)$ of Gleason patterns is initialized as a uniform distribution, assuming equal probability of each possible class.
2) Likelihood: AI predictions are interpreted as likelihoods $P(p_j | g_i, image)$.
3) Posterior Calculation: Using Bayes Theorem, the posterior distribution is calculated as:

$$P(g_i | p_j, image) = \frac{P(p_j | g_i, image) P(g_i | image)}{P(p_j | image)},$$

where $P(p_j | image)$ is the marginal likelihood ensuring the posterior sums to 1.

*Posterior Aggregation.* We aim to leverage an ensemble of $m$ AI models $\{f_1, f_2, \ldots, f_m\}$ each trained to predict the Gleason pattern distribution for each pixel $p_j$. Assuming the AI models provide independent and complementary information, each subsequent model, $f_k$ for $k = 2,3, \ldots, m$, can be integrated by repeating step three above with the posterior distribution $P(g_i | p_j, image)$ from $f_{k-1}$ as the prior distribution.

*AI Model Training.* Here, one AI model {f1} was trained on the PANDA dataset. Details of the model architecture and training process can be found in [11]. Here, we changed the inference module to report probabilities per output class $g_i$ instead of the argmax from the original implementation.

*Gleason Scoring.* The final Gleason grade is derived by 1) Assigning a Gleason pattern to each pixel based on the maximum a posteriori estimates of $P(g_i | p_j, image)$, and 2) Calculating the overall Gleason score as the sum of the primary and secondary pattern scores per standard clinical practice [4].

*Metrics*. The total accuracy and Prevalence Adjusted Bias-Adjusted Kappa (PABAK) metrics, used to compare different annotations, are described elsewhere [11]. In addition to these metrics, we use Shannon entropy [17] to analyze the distributions of probabilities for each pixel.

*Annotation Comparison.* We evaluated the performance of this single AI model on two annotation sets: 1) Expert pathologist annotations; 2) Student annotator labels. We also compared each of these sets of annotations to the PANDA labels.

*Panel Predictions*. To assess the applicability of panel predictions, we selected one random image from the test set and examined the entropy $H = - \sum_k p_k \log(p_k)$ (where $p_k$ is the probability of each class) of the probability distributions for each pixel.

**Results & Discussion**

*Hand Curations*. We compiled a dataset of 81 hand annotated whole slide images. This process reflects inter-annotation variation as 51 individuals contributed to the annotation process. In this work the dataset is used as a test set, but as the dataset grows, it will serve as a crucial resource for training other AI models to enrich the Bayesian framework.

*Accuracy Assessment.* Table 1 shows the accuracy and PABAK scores for the pathologist and student annotations. Compared to the PANDA dataset, the model achieves an average score of 0.81, similar to intra-pathologist comparisons [11] as a result of being trained on this data. The low accuracy between hand annotations of pathologists and students alike when compared to the PANDA dataset suggests the quality difference between our expert annotations and the available PANDA labels. In the future, we anticipate models trained on expert annotations to outperform the current model in clinical applications.

|  | Overall Accuracy | PABAK |
|---|---|---|
| Pathologist vs. PANDA | 0.29 | -0.42 |
| Pathologist vs. Model | 0.27 | -0.45 |
| Model vs. PANDA (on 51 images with pathologist annotations) | 0.82 | 0.64 |
| Students vs. PANDA | 0.30 | -0.40 |
| Students vs. Model | 0.26 | -0.48 |
| Model vs. PANDA (on 81 images with student annotations) | 0.8 | 0.61 |

**Table 1**. Accuracy and PABAK values for the images annotated by pathologists and the images annotated by students

*Assessment of Panel Predictions.* Figure 2 shows how pixel-level prediction maps allow the pathologist to zoom in on certain regions of the image and get a more thorough understanding of the model's prediction. The probability distribution mimics the inter- and intra- pathologist variability present in real world data.

It is akin to asking a panel of expert pathologists about their respective grading on a given image – the preferred 2nd and 3rd opinion-seeking approach of many cancer patients [18].

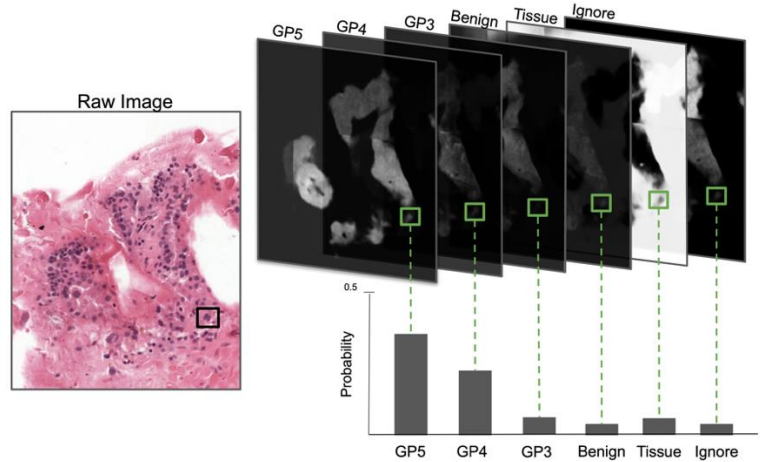

**Figure 2**. *1* For each pixel in a whole slide image, a posterior distribution is predicted as panel opinion, and presented as a probability distribution for each individual class.

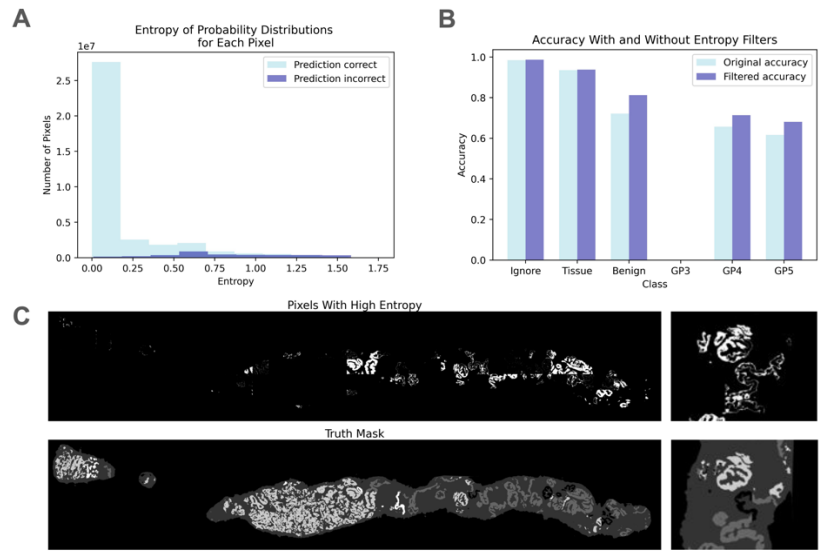

**Figure 3**. A) Pixel-based entropy distributions for the predicted probabilities. B) Accuracy before and after entropy threshold. C) Mapping of pixels with higher entropy (top) compared to the truth mask (bottom). On the right, regions of interest demonstrating how probability distributions can focus pathologists' attention.

We investigated the usability of a distribution as an estimate of the expert panel by using one randomly sampled image from the test set. In this image, pixels predicted incorrectly had a higher entropy on average than those predicted correctly (Figure 3A), indicating that pathologists can be more confident in predictions when the probability is concentrated on one class. Moreover, by excluding pixels above an arbitrarily chosen entropy of 1.2, the model's accuracy on the remaining pixels increases. Here, 72% of the pixels that have a true benign label were correctly classified by the model. If high entropy pixels are removed, accuracy increases to 81%. Similarly, every other class present in the image also increased in accuracy given this threshold (Figure 3B), indicating these pixels are areas needing close attention from pathologists. Plotting these filtered pixels gives pathologists a map of where to focus their time (Figure 3C).

While these results only represent the evaluation of a single example image in the test set, they demonstrate the advantages of training models to predict class probabilities rather than a single prediction. This method shows that these probability distributions can be used to direct the pathologist's attention to the areas of the image that are most likely to be classified incorrectly by the model and thus reduce the error of predictions.

**Conclusion**

We presented a new hand-curated prostate cancer dataset and a novel Bayesian framework that address key challenges hindering the clinical adoption of AI for prostate cancer detection and grading. A key strength of this framework lies in its adaptability and ability to seamlessly integrate new AI models and annotator panels as they become available. This adaptability ensures that the framework can continually refine its predictions, and incrementally improve its performance and generalizability as more diverse data sources are incorporated. Also, using model output as pixel-based probability distributions rather than a single label can reduce prediction errors and save time as pathologists can be directed to areas where additional time and attention are required.

Ultimately, by presenting methods for overcoming the challenges of annotation variability and dataset limitations, this work presents a promising direction toward AI tools that can consistently and reliably support pathologists in accurate prostate cancer grading. While the current results are compelling, there remains a need for larger multi-center studies to validate the generalizability of our approach across diverse patient populations, imaging protocols, and clinical settings. We plan to continue expanding the current dataset and exploring more advanced techniques (such as prior specification, likelihood modeling, and interpretable uncertainty quantification) to further enhance the framework's performance and clinical applicability. We encourage other institutions and professionals to participate in dataset annotation and the further training of AI models.

**Data and Code Availability**

Annotations and models can be made available upon request. A public dataset release is planned.